\documentclass[final,twocolumn]{elsarticle}

\usepackage{hyperref}
\usepackage{amssymb}
\usepackage{amsfonts}
\usepackage{amsmath}
\usepackage{graphicx}
\usepackage{dcolumn}
\usepackage{bm}
\usepackage{color}
\usepackage{natbib}

\newcommand{\sslash}{\mathbin{/\mkern-4mu/}}

\journal{Physica D}

\graphicspath{{Figures/}}

\begin{document}

\begin{frontmatter}

\title{Non-Conservative Variational Approximation for Nonlinear Schr\"{o}dinger Equations}

\author{J. Rossi, R. Carretero-Gonz\'{a}lez}
\address{Nonlinear Dynamical System Group,%
\footnote{\texttt{URL}: \href{http://nlds.sdsu.edu}{http:$\sslash$nlds.sdsu.edu}}
Computational Science Research Center,%
\footnote{\texttt{URL}: \href{http://www.csrc.sdsu.edu}{http:$\sslash$www.csrc.sdsu.edu}}
and Department of Mathematics and Statistics,
San Diego State University,
San Diego, CA 92182-7720, USA}

\author{P.G. Kevrekidis}
\address{Department of Mathematics and Statistics,
University of Massachusetts,
Amherst, MA 01003-4515, USA}


\begin{abstract}
Recently, Galley~[Phys.~Rev.~Lett.~\textbf{110}, 174301 (2013)] proposed 
an initial value problem formulation of Hamilton's principle applied to 
non-conservative systems.  Here, we explore 
this formulation for complex partial differential equations of the 
nonlinear Schr\"{o}dinger (NLS) type, examining the 
dynamics of the coherent solitary wave structures of such models by means
of a non-conservative variational approximation (NCVA).  We compare the 
formalism of the NCVA to two other variational techniques used in dissipative 
systems; namely, the perturbed variational approximation and a generalization of the 
so-called Kantorovich method.  All three variational techniques produce 
equivalent equations of motion for the perturbed NLS models studied herein.  
We showcase the relevance of the NCVA method by exploring test 
case examples within the NLS setting including combinations of linear and 
density dependent loss and gain. We also present an example applied to 
exciton polariton condensates that intrinsically feature loss and a 
spatially dependent gain term.
\end{abstract}


\end{frontmatter}


\section{Introduction}

Variational methods are commonly used to describe 
the dynamics of nonlinear waves in nonlinear optics, and atomic physics~\cite{Boris:02,Boris:06,Dauxois:03,Kivshar:03}.  These methods rely on a well-informed ansatz substituted in the Lagrangian or Hamiltonian formulation of a 
complex, infinite dimensional system. 
This ansatz reduces an original partial differential equation (PDE) model to a few
degrees of freedom establishing equations describing the approximate dynamics appropriately projected into the solution space spanned by the ansatz.
The variational approximation (VA) method projects the high-dimensional dynamics to a low-dimensional dynamical system for 
the time-dependent parameters that encapsulate 
the qualitative and quantitative behavior of the original complex system.
It is important to note that there is an intrinsic drawback of VA methods in that
they have a strong restriction when projecting the infinite-dimensional dynamics
of the original PDE to a small finite-dimensional ansatz subspace.
This projection is known to potentially lead to invalid results~\cite{Kaup:96},
a feature which is naturally expected (given the large reduction in the
number of degrees of freedom) when the full PDE dynamics ceases to be 
well-described by the selected ansatz.
Nonetheless, there have been some efforts to control the corrections of the
VA to increase the accuracy of the results~\cite{Kaup:07}.
Fundamentally, the variational method relies on the existence of a Lagrangian or Hamiltonian structure from which the Euler-Lagrange equations can be derived.  This prerequisite limits the application of the variational approach to conservative, \textit{closed  systems}.

The recent work by Galley~\cite{ref1,Galley:14} offers a new perspective to the
classical mechanical formulations by recognizing that the Hamilton-Lagrangian
formulation has the key feature 
of being a boundary value problem in time although it is
used to derive equations of motion that are solved with initial data.
By treating the extremization problem as an initial value problem, variational
calculus can be applied to non-conservative systems.  Although Galley's proposal
was for classical mechanical, few-degree-of-freedom systems, i.e., systems described by ordinary differential
equations (ODEs), it paved the way for its application to dispersive nonlinear PDEs.
The relevance and validity of the extension of Galley's method to nonlinear PDEs was 
showcased in the recent work of Ref.~\cite{ref4} where examples cases based on a
$\mathcal{PT}$-symmetric sine-Gordon and $\phi^4$ models were presented.
This work recognized that not only can the method be utilized for
infinite degree-of-freedom systems, but its Lagrangian underpinning
enables a variational approach to be developed
for non-conservative PDE systems. Ref.~\cite{ref4}
illustrated 
very good agreement between the original PDE dynamics and the reduced ODE 
model, obtained from the corresponding non-conservation variational
approximation (NCVA).

In this manuscript we present the extension of this NCVA method for the
specific, but broadly important/applicable 
case of the nonlinear Schr\"odinger (NLS) equation.
The latter model and its variants are of principal interest
to applications from optical physics~\cite{Kivshar:03}, 
atomic physics~\cite{pitas} and other areas of mathematical
physics~\cite{chap01:ablowitz}, not only in their conservative,
but also in dissipative variants of the model~\cite{Aranson:02}.
The details of our presentation are as follows.
Section~\ref{sec:NC} presents the variational formalism.
For completeness, the section starts with a short review on
Galley's method as originally introduced for finite-dimensional
Hamiltonian systems and extended to complex non-conservative
forces. Then, we extend this methodology to the
NLS equation and compare it to the perturbative variational approximations,
examined previously.
We, in fact, prove that these different methods, within the NLS setting, are
equivalent. In Sec.~\ref{sec:Aps} we present a few examples
of NLS settings including non-Hamiltonian terms. We use a combination
of linear and density dependent loss and gain and showcase an
example in the real of polariton condensates that are
out-of-equilibrium as they intrinsically contain loss and gain
terms. Finally, in Sec.~\ref{sec:conclu} we present our conclusions
and discuss a few avenues for further exploration.

\section{Non-conservative Variational Approximation Formalism}
\label{sec:NC}

\subsection{Non-conservative Approximation for Finite-Dimensional Systems}
\label{sec:NC_Ham}

Let us discuss the non-conservative variational approach (NCVA),
based on the non-conservative variational formulation
introduced by Galley in Ref.~\cite{ref1}. Our aim here will be
generalize the latter to include complex variables and
complex non-conservative forces. Subsequently, we will use
the NCVA to study the dynamical characteristics of nonlinear
waves in the perturbed NLS model.
In Ref.~\cite{ref1}, Galley highlights that the time-symmetric and conservative dynamics in
Hamiltonian systems is due to the boundary value form of the action 
extremization problem.
As such, the original formalism based on extremization of an action functional is
predicated on the conservative nature of the system. Therefore, a direct application
of action extremization when the system includes non-conservative terms is, by
construction, not viable.
As Galley points out, in simple classical mechanics cases where the dissipation 
forces are local in time and {\em linear} in the velocities, it is possible
to employ Rayleigh's dissipation function~\cite{Goldstein}.
However, this formulation cannot be applied to systems with more general 
dissipative forces such as nonlocal or {\em nonlinear} ones.

Therefore, in order to treat systems with general forces, 
Galley proposed to consider the extremization problem
as an {\em initial value problem} instead. The method is based on considering two sets
of variables, $\vec{q}_1$ and $\vec{q}_2$, and applying variational calculus for the
non-conservative system provided $\vec{q}_2 = \vec{q}_1$ after the variation.
Consider the path $\vec{q}(t)$ passing through the (fixed) 
boundary values
$\vec{q}_i$ at $t=t_i$ (initial) and $\vec{q}_f$ at $t=t_f$ (final). 
Suppose that the
system trajectories are described by the following set of $N$ generalized coordinates
and velocities: $\vec{q} \equiv \{q_i\}_{i=1}^N$ and $\dot{\vec{q}} \equiv  \{ \dot{\vec{q}}_i \}_{i=1}^N$. Let us now double both sets of quantities,
$\vec{q} \rightarrow (\vec{q_1}, \vec{q_2})$ and
$\dot{\vec{q}} \rightarrow (\dot{\vec{q}}_1, \dot{\vec{q}}_2)$, and parametrize
both coordinate paths:
\begin{equation}
\vec{q}_{1,2} (t, \epsilon) = \vec{q}_{1,2} (t, 0) + \epsilon\, \vec{\eta}_{1,2}(t),
\end{equation}
where $\vec{q}_{1,2} (t, 0)$ are the coordinates of two stationary paths ($\epsilon \ll 1$)
and $\vec{\eta}_{1,2}(t) $ are arbitrary virtual displacements.
The following equality conditions are required for varying the action such that
the endpoints coincide:
$\vec{\eta}_{1,2}(t_i) = 0$, $\vec{q}_{1} (t_f, \epsilon) = \vec{q}_{2} (t_f, \epsilon)$,
and $\dot{\vec{q}}_{1} (t_f, \epsilon) = \dot{\vec{q}}_{2} (t_f, \epsilon).$
%
Therefore, the equality condition does not fix either value at the final time.  After all variations are performed, both paths are set equal and identified with the physical path, $\vec{q}(t)$, the so-called physical limit (PL).

The total action functional for $\vec{q}_1$ and $\vec{q}_2$ is defined as the total
line integral of the Lagrangian along both paths plus the line integral of a
function $\mathcal{R}$ which describes the generalized non-conservative forces and
depends on both paths $\{\vec{q}_n\}_{n=1}^2$:
\begin{eqnarray}
S\left[\vec{q}_n\right] &\equiv & \int_{t_i}^{t_f} dt \mathcal{L} (\vec{q}_1, \dot{\vec{q}}_1) - \int_{t_f}^{t_i} dt \mathcal{L}(\vec{q}_2, \dot{\vec{q}}_2)
\nonumber
 \\[1.0ex]
 &&+ \int_{t_i}^{t_f} dt \mathcal{R} (\vec{q}_n, \dot{\vec{q}}_n, t ), \nonumber \\[2.0ex]
&=& \int_{t_i}^{t_f} dt \left[\mathcal{L} (\vec{q}_1, \dot{\vec{q}}_1)  - \mathcal{L}(\vec{q}_2, \dot{\vec{q}}_2) + \mathcal{R} (\vec{q}_n, \dot{\vec{q}}_n, t) \right]\quad.
\end{eqnarray}
The above action defines a new Lagrangian:
\begin{equation}
\Lambda (\vec{q}_n, \dot{\vec{q}}_n) \equiv  \mathcal{L} (\vec{q}_1, \dot{\vec{q}}_1) - \mathcal{L }(\vec{q}_2, \dot{\vec{q}}_2) + \mathcal{R} (\vec{q}_n, \dot{\vec{q}}_n, t).
\label{eq:action}
\end{equation}
If $\mathcal{R}$ is written as the difference of two potentials $V(\vec{q}_1) - V(\vec{q}_2)$,
i.e., under the presence of conservative forces,
then it is absorbed into the difference of two (suitably redefined)
Lagrangians, leaving $\mathcal{R}$ effectively to be zero.
On the other hand, a nonzero (nontrivial) 
$\mathcal{R}$ corresponds to {\em non-conservative} forces and couples
the two paths together.

For convenience, following Ref.~\cite{ref1}, 
we make a change of variables to $\vec{q}_+ = (\vec{q}_1 +\vec{q}_2)/2$
and $\vec{q}_- = \vec{q}_1  -\vec{q}_2$. In this way, $\vec{q}_- \rightarrow 0$
and  $\vec{q}_+ \rightarrow \vec{q}$ in the physical limit.
The conjugate momenta are then $\vec{q}_\pm = \partial \Lambda / \partial \dot{\vec{q}}_\mp$
and the paths are parametrized by
$\vec{q}_{\pm} (t, \epsilon) = \vec{q}_{\pm} (t, 0) + \epsilon \vec{\eta}_{\pm}(t)$.
Therefore, the new action is stationary under these variations if
$|dS[\vec{q}_\pm]/d\epsilon|_{\epsilon = 0} = 0$ for all $\vec{\eta}_{\pm}$.


In the original $\vec{q}_{1,2}$ coordinates the equations of motion may be written,
using the Euler-Lagrange equations, as
$d \vec{\pi}_{1,2}/dt  = \partial \Lambda / \partial \vec{q}_{1,2}$ with
$\vec{\pi}_{1,2} = (-1)^{1,2} \partial \Lambda / \partial \dot{\vec{q}}_{1,2}$.
%
Similarly, in the $\vec{q}_\pm$, the Euler-Lagrange equations yield to the
equations of motion. However, in the physical limit, only the
$ \partial \Lambda/\partial \vec{q}_- = d \vec{\pi}_+ /dt$ equation survives,
such that the trajectory is defined by
\begin{equation}
\label{EL1}
\frac{d}{dt} \vec{\pi} (\vec{q},\dot{\vec{q}}) = \Bigg[ \frac{\partial \Lambda}{\partial \vec{q}_-} \Bigg]_{\rm PL} = \frac{\partial \mathcal{L}}{\partial \vec{q}} + \Bigg[ \frac{\partial \mathcal{R} }{\partial \vec{q}_-}\Bigg]_{\rm PL},
\end{equation}
with conjugate momenta
\begin{equation}
\label{EL2}
\vec{\pi} (\vec{q},\dot{\vec{q}}) = \left[ \frac{\partial \Lambda}{\partial \dot{\vec{q}}_-} \right]_{\rm PL} = \frac{\partial \mathcal{L}}{\partial \dot{\vec{q}}} + \left[ \frac{\partial \mathcal{R}}{\partial \dot{\vec{q}}_-}\right]_{\rm PL}.
\end{equation}
In the presence of conservative forces (i.e., $\mathcal{R} = 0$), the usual Euler-Lagrange
equations are recovered.  However, for non-conservative forces, a nonzero $\mathcal{R}$
modifies the trajectories as per Eqs.~(\ref{EL1}) and (\ref{EL2}). Since in our
considerations below we are concerned with complex non-conservative forces, the
non-conservative term $\mathcal{R}$ will be complex as well.
%
%
The terms in $\mathcal{R}$
are precisely responsible for coupling both the $\vec{q}_1$
and $\vec{q}_2$ paths to each other.
It is important to note that the complex conjugate of the functional terms
in $\mathcal{L}$ are necessary for solving the Euler-Lagrange equations when considering
coordinates containing complex terms ---which is precisely the case for the NLS
under consideration.
In the physical limit, only the Euler-Lagrange equation for the $+$ variables survives.
Therefore, expanding the action in powers of $\vec{q}_-$ the equations of motion
follow the variational principle:
\begin{equation}
\Bigg[\frac{ \delta S[\vec{q}_{\pm}] }{\delta \vec{q}_- (t)}\Bigg]_{\rm PL} = 0.
\end{equation}
This also has the consequence 
that only terms in the new action that are perturbatively linear in
$\vec{q}_-$ contribute to physical forces.

\subsection{Non-Conservative Variational Formulation for Nonlinear Schr\"{o}dinger Equation}
\label{subsec:NC_NLS}

Let us now extend the NCVA formalism for the NLS equation.
It is worth mentioning at this stage that the NLS equation 
is, arguably, a prototypical nonlinear
PDE that supports travelling envelope waves. Namely, the NLS equation is the
normal form for envelope waves~\cite{chap01:sulem}. As such, the NLS equation is a paradigm
for wave propagation and a universal model describing the evolution of complex field
envelopes in nonlinear dispersive media~\cite{dodd,NLPDE}. The NLS equation
models a wide range of physical phenomena including
hydrodynamic waves on deep water \cite{Newell:67},
nonlinear optical systems \cite{Hasegawa:73,has1,ab1,has2,Kivshar:03},
heat pulses in solids \cite{Tappert:70},
plasma waves \cite{Ichikawa:72,Ichikawa:73,infeld},
and
matter waves \cite{BECBOOK,review};
while it has also attracted much interest
mathematically~\cite{chap01:sulem,chap01:ablowitz,chap01:bourgain,segur,zakhbook,acnewell}.

The one-dimensional (1D) NLS equation can be cast in non-dimensional form as~\cite{ref10}
\begin{equation}
i u_t + \frac{1}{2} u_{xx} + g |u|^2 u  = 0,
\label{eq:conservativeNLS}
\end{equation}
where $u(x,t)$ is the complex field and $g$ is the nonlinearity coefficient
[$g=+1$ ($g=-1$) corresponds to an attractive (repulsive) or focusing (defocusing)
nonlinearity].
This form of NLS corresponds to a Hamiltonian PDE.
The Lagrangian density for this conservative NLS is~\cite{ref5, ref6, ref10}
\begin{equation}
\mathcal{L} = \frac{i}{2} (u^* u_t - u u_t^*) + \frac{1}{2} |u_x|^2 - \frac{1}{2} g |u|^4,
\label{eq:conLagragian}
\end{equation}
where $(\cdot)^*$ denotes complex conjugation.
For consistency of notation we will use calligraphic symbols (cf.~$\mathcal{L}$)
to denote densities while their effective (integrated over all $x$) quantities
we will use standard symbols. Namely ${L} = \int_{-\infty}^{\infty} {\mathcal{L}}\,dx$.
In the presence of non-conservative terms ($\mathcal{P}$) that may depend of the
field $u$, its derivatives, and/or its complex conjugate, the NLS takes
the general form:
%
\begin{equation}
i u_t + \frac{1}{2} u_{xx} + g |u|^2 u  =  \mathcal{P}.
\label{eq:nonconservativeNLS}
\end{equation}
Following Ref.~\cite{ref1}, two coordinates are introduced: $u_1$ and $u_2$.
In analogy with the finite-dimensional case of the previous section,
we construct the corresponding total Lagrangian for these two coordinates:
\begin{equation}
\mathcal{L}_T = \mathcal{L}_1 - \mathcal{L}_2 + \mathcal{R},
\label{eq:L12}
\end{equation}
where $\mathcal{L}_i \equiv \mathcal{L}(u_i, u_{i,t}, u_{i,x},..,t)$, for $i = 1,2$,
represents the corresponding conservative Lagrangian
densities for $u_1$ and $u_2$ as defined in Eq.~(\ref{eq:conLagragian}) and
$\mathcal{R}$ contains {\em all} non-conservative terms that originate from the term $\mathcal{P}$
in Eq.~(\ref{eq:conservativeNLS}).
Therefore, by construction, the non-conservative part
of the Lagrangian~(\ref{eq:L12}) must be related to the perturbation term $\mathcal{P}$
in the NLS~(\ref{eq:nonconservativeNLS}) by
\begin{equation}
\mathcal{P}= \left[ \frac{\partial \mathcal{R}}{\partial u_-^*} \right]_{\rm PL},
\end{equation}
such that $\mathcal{R} = \mathcal{P}\, u_-^* + {\rm const}$, where the
constant of integration is with respect to $u_-^*$.
As before, for convenience, $u_+ = (u_1 + u_2)/2$
and $u_- = u_1 - u_2$ are defined in such a way that at the PL
$u_+ \,  \rightarrow \, u$ and $u_- \, \rightarrow \, 0$.  The corresponding conjugate momenta are defined as in Sec.~\ref{sec:NC_Ham} and thus, the equation of motion reads
\begin{equation}
\frac{\partial }{\partial t} \frac{\delta \mathcal{L}}{\delta u_t^*}=  \frac{\delta \mathcal{L}}{\delta u^*} + \left[ \frac{\delta \mathcal{R}}{\delta u_-^* }\right]_{\rm PL},
\end{equation}
where $\delta$ denotes Fr\'{e}chet derivatives.
%
%
%
%

Through this method we recover the Euler-Lagrange equation for the conservative terms
and all non-conservative terms are lumped
into $[ \delta \mathcal{R}/\delta u_-^* ]_{\rm PL}$.
It is crucial to construct an $\mathcal{R}$ such that its derivative with respect to the difference
variable $u_-^* = u_1^* -u_2^*$ at the physical limit gives back the non-conservative or generalized
forces [$\mathcal{P}$ in Eq.~(\ref{eq:nonconservativeNLS})].  A similar variational formulation can be applied to other PDE (or ODE)
models of interest.

\subsection{Perturbed Variational Approach Formalism and Equivalence Proof}

Let us now compare the above methodology using the NCVA to the standard
perturbed variational approach~\cite{nonlinsc}.
Let us consider the non-conservative modified NLS equation Eq.~(\ref{eq:nonconservativeNLS})
with the non-conservative generalized force $\mathcal{P} = \epsilon Q$, where $\epsilon$ is
a formal, small, perturbation parameter ($|\epsilon| \ll 1$).
In this manner, when $\epsilon = 0$, one recovers the conservative
Lagrangian~(\ref{eq:conLagragian}) with corresponding Euler-Lagrange equations:
\begin{equation}
\frac{\partial \bar{{L}}}{\partial \vec{p}} - \frac{d}{dt} \frac{\partial \bar{{L}}}{\partial \dot{\vec{p}}} = 0,
\end{equation}
where $\bar{{L}}(\vec{p}) =  \int \bar{\mathcal{L}}dx$ and
$\bar{\mathcal{L}} \equiv \mathcal{L}[\bar{u}(x,t, \vec{p})]$ is the
conservative Lagrangian (\ref{eq:conLagragian}) evaluated on the chosen
variational ansatz $\bar{u}$ bearing a vector of
variational parameters $\vec{p}$, on which the effective Lagrangian
$\bar{\mathcal{L}}$ depends.
For consistency we will now use a bar over quantities that are evaluated
at the variational ansatz.
To adjust for the presence of small non-conservative terms one needs to find
the remainder of
\begin{equation}
 \frac{\partial \bar{{L}}_T}{\partial \vec{p}} - 
\frac{d}{dt} \frac{\partial \bar{{L}}_T}{\partial \dot{\vec{p}}},
 \label{eq:nonEL}
\end{equation}
which is nonzero, where the Lagrangian $\bar{{L}}_T = \bar{{L}} + \bar{{L}}_\epsilon$
conservative terms $\bar{{L}}$ and the non-conservative terms $\bar{{L}}_\epsilon$.

Applying the (linear) perturbed variational approximation~\cite{nonlinsc,Boris:02}
yields, after obtaining the effective Lagrangian from the Lagrangian density,
the following perturbed Euler-Lagrange equation:
\begin{equation}
 \frac{d}{dt} \frac{\partial \bar{L}}{\partial \dot{p}} - \frac{\partial\bar{ L}}{\partial p}  =
 \epsilon\int_{-\infty}^{\infty} \left( Q^*\frac{\partial \bar{u}}{\partial p}  + Q \frac{\partial \bar{u}^*}{\partial p} \right)dx.
\label{eq:pvaRHS}
\end{equation}
The right hand side above 
is also equivalent to the result of the modified Kantorovich
method~\cite{ref3} yielding in the right hand side of Eq.~(\ref{eq:pvaRHS})
\begin{equation}
\int_{-\infty}^{\infty} \left( \bar{\mathcal{P}}^*\frac{\partial \bar{u}}{\partial p}  + \bar{\mathcal{P}} \frac{\partial \bar{u}^*}{\partial p} \right)dx \equiv 2 \mathrm{Re} \int_{-\infty}^{\infty} \bar{\mathcal{P} }\frac{\partial \bar{u}^*}{\partial p} dx.
 \end{equation}


Let us now establish the equivalence between the above perturbed variational
formulation and the NCVA method.
Given the non-conservative NLS~(\ref{eq:nonconservativeNLS}), where $\mathcal{P}$ is
assumed complex, let us construct the functional $\mathcal{R}$ evaluated 
at the variational ansatz such that
\begin{equation}
\bar{\mathcal{P}} = \left[ \frac{\partial \bar{\mathcal{R}}}{\partial {\bar{u}}_-^*} \right]_{\rm PL}.
\label{eq:criteria1}
\end{equation}
Let us now require that for the evolution of the ansatz the variational parameters 
are real. Thus, the solution which satisfies Eq.~(\ref{eq:criteria1})
and ensures real values for the parameters is:
%
\begin{equation}
\bar{\mathcal{R}} =
\bar{\mathcal{P}}(\bar{u}_{\pm}, \bar{u}_{\pm}^*,\bar{u}_{\pm,t},\ldots) \; 
\bar{u}_-^*  +  c.c.,
\end{equation}
where $c.c.$~stands for complex conjugate.
For ease of notation let us denote by $p$ a single variational parameter (i.e., an
entry of $\vec{p}$) and remind the reader that equations with the symbol $p$
denote a set of couple equations for each of the entries $p$ in $\vec{p}$.
As in Galley's prescription, for simplicity when evaluating in the physical limit,
define the $\pm$ coordinates for each parameter: $p_+ = (p_1 + p_2)/2$ and 
$p_- = (p_1 - p_2)$ such that $p_- = p_-^*$, then we can show the NCVA is 
equivalent to the perturbed variational approximation:
\begin{eqnarray}
 \bar{P}  &=& \int_{-\infty}^{+\infty} \bar{\mathcal{P}} dx = \int_{-\infty}^{\infty} \left[\frac{\partial \bar{\mathcal{R}}}{\partial \bar{u}_-^* } \right]_{\rm PL} dx,
 \end{eqnarray}
projected into the ansatz, such that
 \begin{eqnarray}
\bar{P} &=& \int_{-\infty}^{\infty}  \left[\frac{\partial}{\partial p_-^*} \left(\bar{\mathcal{P}}   \bar{u}_-^* \right)  +
          \frac{\partial}{\partial p_-^*} \left(\bar{\mathcal{P}}^* \bar{u}_-   \right) \right]_{\rm PL} dx, \nonumber  \\
          &=& \int_{-\infty}^{\infty} \left[\bar{\mathcal{P}} \frac{\partial \bar{u}_-^* }{\partial p_-^*} + \bar{u}_-^* \frac{\partial \bar{\mathcal{P}}  }{\partial p_-^*}  + \bar{\mathcal{P^*}} \frac{\partial \bar{u}_- }{\partial p_-^*} + \bar{u}_- \frac{\partial \bar{\mathcal{P^*}}  }{\partial p_-^*}    \right]_{\rm PL} dx, \nonumber \\
&=& \int_{-\infty}^{\infty} \left(
\bar{\mathcal{P}}^* \frac{\partial  \bar{u}}{\partial p^*}
+
\bar{\mathcal{P}} \frac{\partial  \bar{u}^*}{\partial p }
\right) dx, \label{eq:NCVAequiv}
\end{eqnarray}
since $[\bar{u}_-^*]_{\rm PL} = [\bar{u}_-]_{\rm PL} = 0$.
The non-conservative integral in the Euler-Lagrange equation derived in Eq.~(\ref{eq:NCVAequiv}) is equivalent to the perturbed variational approximation in Eq.~(\ref{eq:pvaRHS}).
%

\section{Applications of the NCVA to the NLS}
\label{sec:Aps}

\subsection{Linear loss}

As a first example for the application of the NCVA, we use the focusing ($g=+1$)
NLS equation with a linear loss term of strength $\epsilon$:
\begin{equation}
iu_t + \frac{1}{2} u_{xx} + |u|^2 u = -i  \epsilon  u.
\label{eq:NLSLL}
\end{equation}
In the absence of the linear loss ($\epsilon=0$), the NLS~(\ref{eq:NLSLL})
admits the well-known, bright, solitary wave solutions~\cite{BECBOOK,nonlinsc}.
Therefore, in order to follow the effects of the linear loss on the
soliton, we choose a bright soliton ansatz with arbitrary height $a$,
inverse width $w$, center position $\xi$, speed $c$, chirp $b$, and phase $\phi$
as follows:
\begin{equation}
u_A(x,t;\vec{p}) = a\, \mathrm{sech}(w\, (x-\xi)) e^{i(b\, (x-\xi)^2 + c\,(x-\xi)+\phi)},
\label{ansatz1}
\end{equation}
where the vector of time-dependent parameters corresponds to: $\vec{p}=(a,w,\xi,c,b,\phi)$.
For the conservative variational approximation, one can define an effective Lagrangian
$\bar{L}$, with the expected Euler-Lagrange equations of motion.

In the NCVA framework, the $\bar{u}_1$ and $\bar{u}_2$ ans{\"a}tze
are defined as in Eq.~(\ref{ansatz1}):
\begin{eqnarray}
\bar{u}_1 &=&u_A(x,t;\vec{p}_1),
\label{eq:psi1} \\
\bar{u}_2 &=&u_A(x,t;\vec{p}_2),
\label{eq:psi2}
\end{eqnarray}
where each solution has its corresponding parameters:
$\vec{p}_1=(a_1,w_1,\xi_1,c_1,b_1,\phi_1)$ and
$\vec{p}_2=(a_2,w_2,\xi_2,c_2,b_2,\phi_2)$.
According to the non-conservative variational method the Lagrangian
is $\mathcal{L}_T = \mathcal{L}_1 - \mathcal{L}_2 + \mathcal{R}$ where

\begin{eqnarray}
\bar{\mathcal{L}_1} &=& \frac{i}{2} \left(\bar{u}_1 
\bar{u}_{1,t}^* - \bar{u}_1^* \bar{u}_{1,t}\right) + 
\frac{1}{2} |\bar{u}_{1,x}|^2 - \frac{1}{2}|\bar{u}_1|^4,  \quad\\[1.0ex]
\bar{\mathcal{L}}_2 &=& \frac{i}{2} \left(\bar{u}_2 \bar{u}_{2,t}^* - \bar{u}_2^* \bar{u}_{2,t}\right) + \frac{1}{2} |\bar{u}_{2,x}|^2 - \frac{1}{2}|\bar{u}_2|^4, \quad\\[1.0ex]
\bar{\mathcal{R}} &=&  i \epsilon (\bar{u}_2\bar{u}_1^* - \bar{u}_1 \bar{u}_2^* ).
\end{eqnarray}
%
We then write the effective Lagrangian
$\bar{L} = \int_{-\infty}^{\infty}\bar{\mathcal{L}}_Tdx = \int_{-\infty}^{\infty} \bar{\mathcal{L}}_1dx - \int_{-\infty}^{\infty} \bar{\mathcal{L}}_2 dx + \int_{-\infty}^{\infty} \bar{\mathcal{R}} dx$, for which
$\bar{\mathcal{L}}_1$ and $\bar{\mathcal{L}}_2$ recover the same equations of motion as the conservative
variational approximation.
%
This yields the modified Euler-Lagrange equations:
\begin{equation}
\frac{\partial \bar{L}}{\partial p} - \frac{d}{dt} \left( \frac{\partial \bar{L} }{\partial\dot{p}}\right) + \int_{-\infty}^{\infty} \left[ \frac{\partial \bar{\mathcal{R}}}{\partial p_-} \right]_{\rm PL} dx = 0.
\end{equation}
After integration and simplification, the effective Lagrangian is given by:
\begin{eqnarray*}
\bar{L} &=&\frac{1}{3}a_1^2 w_1+2 \frac{a_1^2\dot{\phi_1}}{w_1}+\frac{a_1^2 c_1^2}{w_1}-2\frac{a_1^2c_1\dot{\xi_1}}{w_1}-\frac{2}{3}\frac{a_1^4}{w_1}  \nonumber \\
&&+\frac{1}{3}\frac{a_1^2 b_1^2 \pi^2}{w_1^3}+  \frac{1}{6}\frac{a_1^2\dot{b_1} \pi^2}{w_1^3}- \frac{1}{3}a_2^2 w_2 - 2 \frac{a_2^2\dot{\phi_2}}{w_2} - \frac{a_2^2 c_2^2}{w_2} \nonumber \\
&&+2\frac{a_2^2c_2\dot{\xi_2}}{w_2}+\frac{2}{3}\frac{a_2^4}{w_2}  -\frac{1}{3}\frac{a_2^2 b_2^2 \pi^2}{w_2^3}-\frac{1}{6}\frac{a_2^2\dot{b_2} \pi^2}{w_2^3} \nonumber \\
&& + i \epsilon \int_{-\infty}^{\infty} \left[ \frac{\partial}{\partial p_-} \Big( \bar{u}_2\bar{u}_1^* - \bar{u}_1\bar{u}_2^* \Big) \right]_{\rm PL} dx .
\end{eqnarray*}
Although the effective Lagrangian is expressed in 1,2 coordinates (for brevity), the Lagrangian must be expanded into the $\pm$ coordinates in order to evaluate the physical limit.
For all the parameters we made the following $\pm$ coordinate substitutions into the expression for the total (effective) Lagrangian:
\begin{equation}
p_1 = \frac{(2p_+ + p_-)}{2}, \\ \quad \quad
p_2 = \frac{(2p_+ - p_-)}{2},
\end{equation}
with $p_1 \in \{ a_1,  b_1, c_1, d_1, \omega_1, \xi_1\}$ and $p_2 \in \{ a_2, b_2, c_2, d_2, \omega_2, \xi_2\}$.

\begin{figure*}[htb]
\centering
\includegraphics[width=8.6cm]{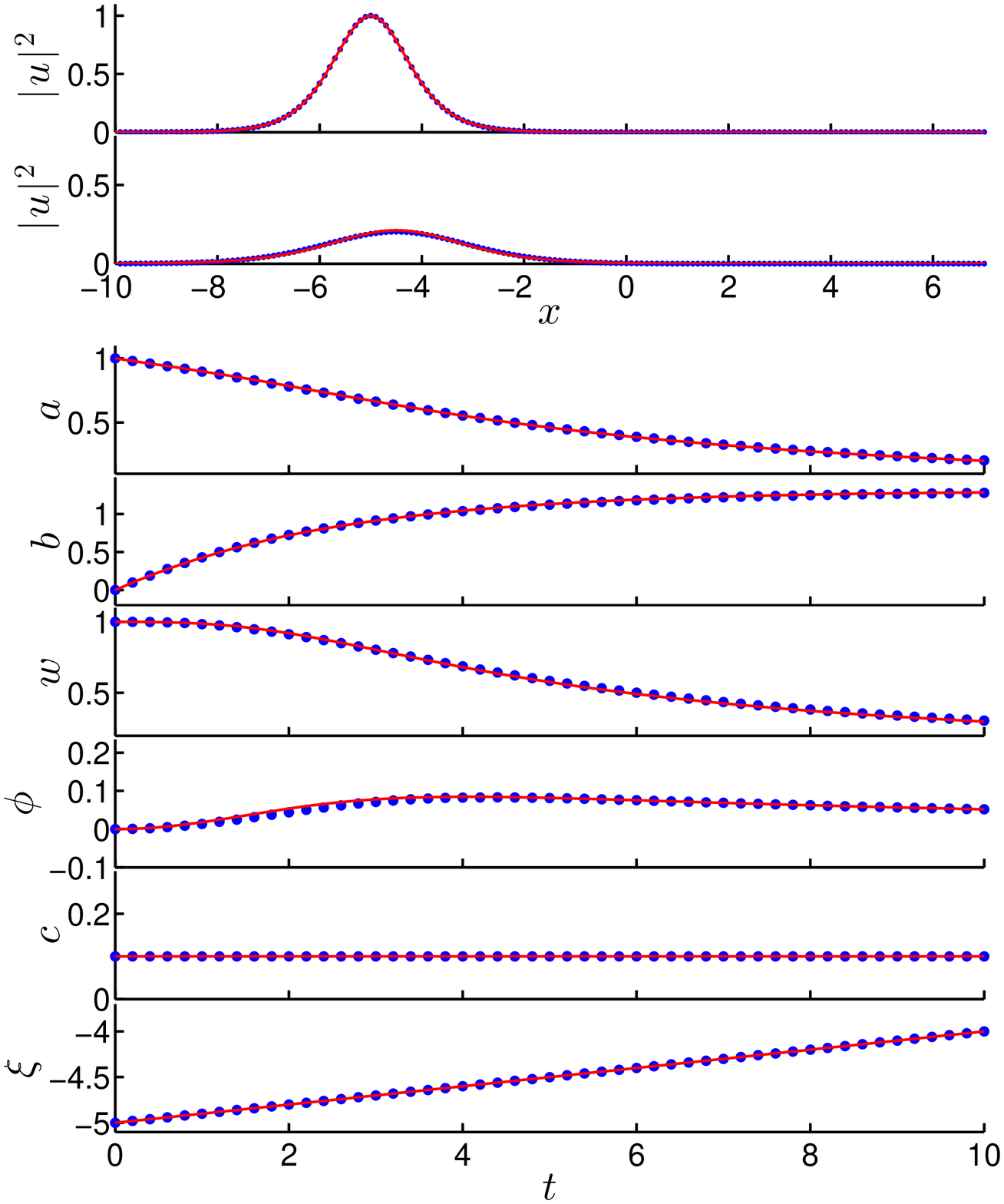}
~
\includegraphics[width=8.6cm]{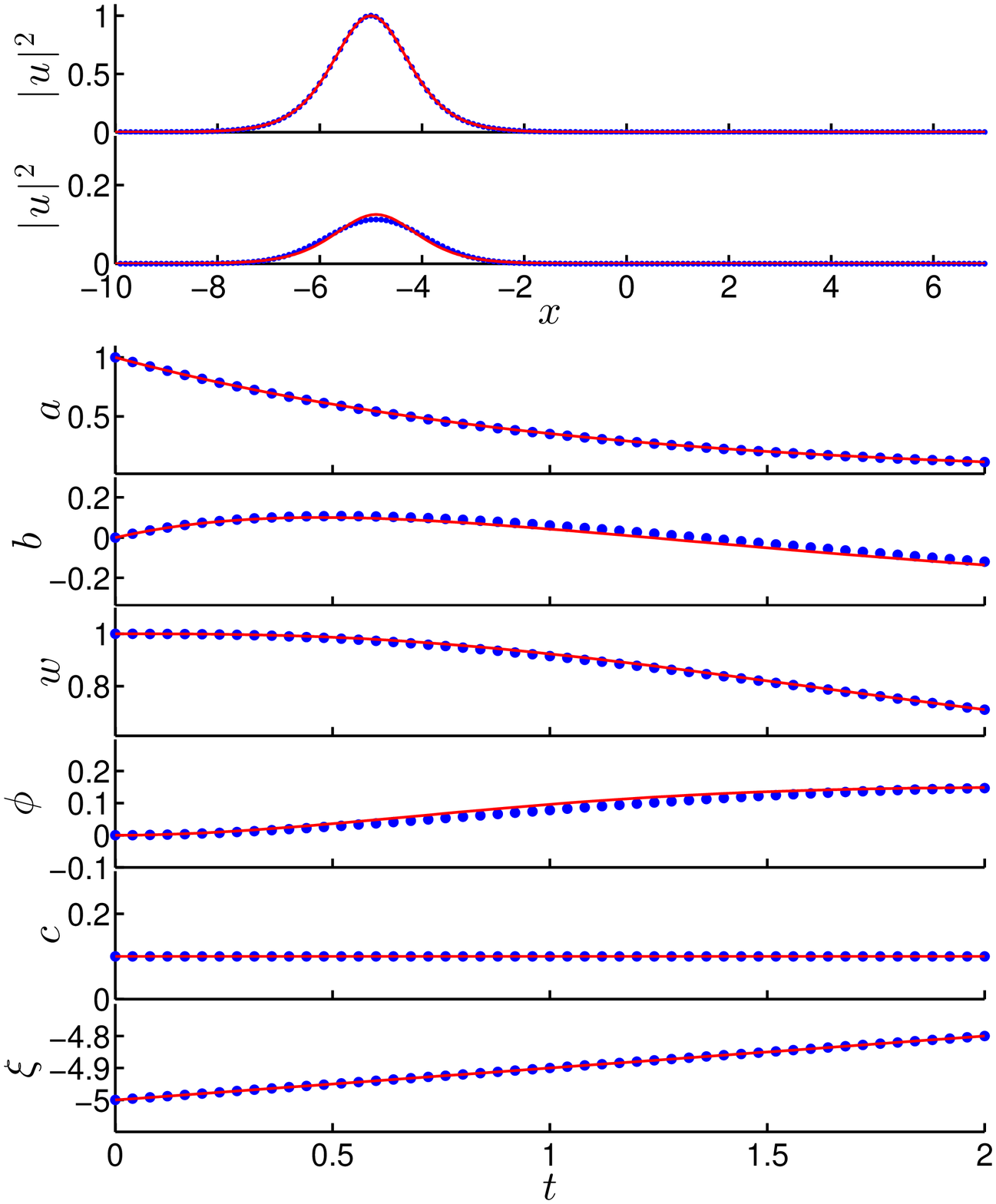}
\vspace{-0.2cm}
\caption{
Evolution of an NLS bright soliton solution under the presence of linear loss
of strength $\epsilon=0.1$ (left) and $\epsilon=1$ (right).
A bright soliton, as described by Eq.~(\ref{ansatz1}), is used as an initial condition
with the parameters:
$a(0)=w(0)=1$, $c(0)=0.1$, $\xi(0) = -5$, and $b(0)=\phi(0)=0$.
Depicted are the comparison of the NCVA approximation of Eq.~(\ref{eq:NCVALL})
(red lines) with the full, numerical, NLS evolution of Eq.~(\ref{eq:NLSLL})
(blue dots).
The top two panels depict the density $|u|^2$ at the initial time (top subpanel)
and at time $t=1/\epsilon$ (second subpanel).
The bottom six subpanels depict the evolution of the NCVA ansatz parameters
$a$, $b$, $c$, $\xi$, $w$, and $\phi$. For the full NLS evolution, 
the parameters are
extracted by projecting the current solution into the NCVA ansatz using least-squares fitting.
For $\epsilon=0.1$ (left) the system is evolved for a total 
time of $t=1/\epsilon$, while for $\epsilon=1$ (right) the final time
is taken to be $t=2/\epsilon$.
\label{fig2}}
\end{figure*}

Therefore, the full expansion with similar terms grouped together is $\bar{L} = \bar{L}_1 - \bar{L}_2 +\bar{R}$ where
\begin{equation}
\bar{R} = \int_{-\infty}^{\infty} \bar{\mathcal{R}} dx =  i\epsilon \int_{-\infty}^{\infty} \left[ \frac{\partial}{\partial p_-} \left(  \bar{u}_2\bar{u}_1^* - \bar{u}_1\bar{u}_2^* \right) \right]_{\rm PL} dx.  \label{barLNClinear}
\end{equation}
$\bar{L}_1$ and $\bar{L}_2$ converge to the physical limit to recover the standard soliton evolution equations, $\bar{L}$, i.e.~the variational approximation for the Hamiltonian, conservative, NLS equation. In this Hamiltonian case
(i.e., in the absence of perturbations), we obtain the following
equations of motion:
\begin{eqnarray} \begin{cases}
\dot{a}  = -ab, \\[1.0ex]
\dot{b}  = \frac{2}{\pi^2}w^4 - \frac{2}{\pi^2}a^2 w^2 - 2 b^2, \\[1.0ex]
\dot{c} = 0 , \\[1.0ex]
\dot{\xi} = c, \\[1.0ex]
\dot{w} = -2bw, \\[1.0ex]
\dot{\phi} = \frac{5}{6} a^2 - \frac{1}{3} w^2 + \frac{1}{2} c^2.
\end{cases} \label{eq:CODEsNLS}\end{eqnarray}
In the presence of the non-conservative term $\bar{R}$, we expand in the $\pm$ coordinate systems 
and find the integrals:
\begin{eqnarray}
\int_{-\infty}^{\infty} \left[ \frac{\partial \bar{\mathcal{ R}}}{\partial a_-} \right]_{\rm PL} dx &=&  0, 
\nonumber  \\
\int_{-\infty}^{\infty} \left[ \frac{\partial  \bar{\mathcal{R}}}{\partial b_-} \right]_{\rm PL} dx &=& -\frac{\pi^2 \epsilon a^2}{3 w^3}, 
\nonumber \\
\int_{-\infty}^{\infty} \left[ \frac{\partial  \bar{\mathcal{R}}}{\partial c_-} \right]_{\rm PL} dx &=& 0, 
\nonumber \\
\int_{-\infty}^{\infty} \left[ \frac{\partial  \bar{\mathcal{R}}}{\partial \xi_-} \right]_{\rm PL} dx &=&  \frac{4 \epsilon a^2 c}{w}, 
\nonumber \\
\int_{-\infty}^{\infty} \left[ \frac{\partial  \bar{\mathcal{R}}}{\partial w_-} \right]_{\rm PL} dx &=& 0, 
\nonumber \\
\int_{-\infty}^{\infty} \left[ \frac{\partial  \bar{\mathcal{R}}}{\partial \phi_-} \right]_{\rm PL} dx &=&  -\frac{4 \epsilon a^2}{w}. 
\nonumber 
\end{eqnarray}
%
Therefore, combining the conservative and non-conservative parts, the
NCVA yields the following equations of motion for the NLS with linear loss:
\begin{equation}
\begin{cases} \dot{a} = - a \epsilon  - a b, \\[1.0ex]
\dot{b}  = \frac{2}{\pi^2}w^4 - \frac{2}{\pi^2}a^2 w^2 - 2 b^2, \\[1.0ex]
\dot{c} = 0 , \\[1.0ex]
\dot{\xi} = c, \\[1.0ex]
\dot{w} = -2bw, \\[1.0ex]
\dot{\phi} = \frac{5}{6} a^2 - \frac{1}{3} w^2 + \frac{1}{2} c^2,
\end{cases}
\label{eq:NCVALL}
\end{equation}
which correspond to the same dynamics as the conservative case (\ref{eq:CODEsNLS}) with the added loss term $-a\epsilon$ for the evolution of the amplitude.


The resulting dynamics comparison between the NCVA ODEs and the numerically integrated NLS
is depicted in 
Fig.~\ref{fig2} for
$\epsilon = 0.1,$ (left set of panels) and $\epsilon = 1$ (right set of panels).  
The figure depicts in the respective top two panels the spatial 
profiles of the densities $|u|^2$ at the initial time ($t=0$) and at a time of
order $1/\epsilon$ for the PDE and ODE solutions and the remaining panels depict 
the evolution of the NCVA parameters.
To compare the NCVA evolution of the parameters to the full NLS numerics, the
numerical NLS solutions are projected (using a least-squares fitting) onto the variational
ansatz $u_A$ at discrete time intervals in order extract the 
variational dynamical parameters associated with the entries of
the vector $\vec{p}$.
The time evolution for these parameters is depicted in the bottom six rows of
panels in the figure.
As observed from the figure, the agreement between the NCVA system of ODEs and the
numerical PDE integration is very good and accurately reflects the
main dynamical features of the soliton solutions, such as the decrease
of its amplitude ($a$), the increase of its width (hence the decrease of
the inverse width controlling parameter $w$) etc.
The high accuracy of the NCVA results, even in the case of large dissipation
(see right set of panels in Fig.~\ref{fig2} where $\epsilon=1$), may be deemed reasonable
to expect in this case of linear dissipation.

\subsection{Density dependent loss}

\begin{figure}[htb]
\centering
\includegraphics[width=8.6cm]{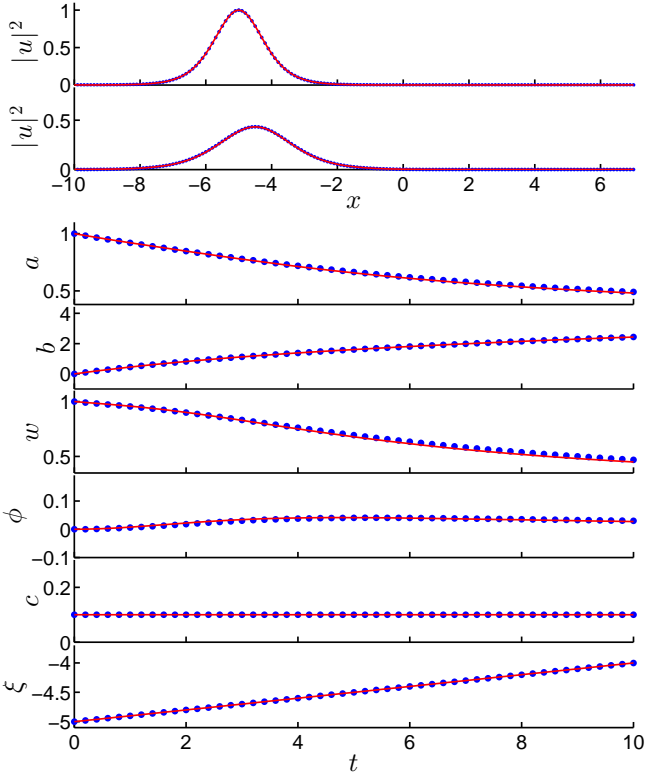}
\vspace{-0.2cm}
\caption{
Evolution of an NLS bright soliton solution under the presence of density dependent loss
of strength $\epsilon=0.1$. The full numerical solution is obtained from
Eq.~(\ref{eq:NLSDD}) (blue dots) while the NCVA results are obtained
from Eq.~(\ref{eq:NVCADD}) (red lines).
Same layout of the panels as in previous figures.
\label{fig3}}
\end{figure}

As a second example, we use the attractive NLS equation with a density dependent (nonlinear)
loss term of strength $\epsilon$:
\begin{equation}
iu_t + \frac{1}{2} u_{xx} + |u|^2 u = -i  \epsilon  |u|^2 u.
\label{eq:NLSDD}
\end{equation}
We follow the same procedure as in the previous example. The conservative terms
are the same and we just need to obtain the non-conservative ones. Taking $\bar{R}$ 
and expanding in the $\pm$ coordinates yields the non-conservative terms:
\begin{eqnarray}
\int_{-\infty}^{\infty} \left[ \frac{\partial  \bar{\mathcal{R}}}{\partial a_-} \right]_{\rm PL} dx &=& 0,  \nonumber  \\
\int_{-\infty}^{\infty} \left[ \frac{\partial  \bar{\mathcal{R}}}{\partial b_-} \right]_{\rm PL} dx &=& -\frac{2\pi^2\epsilon}{9}  \frac{a^4}{w^3} + \frac{4\epsilon}{3} \frac{a^4}{w^3}, \nonumber \\
\int_{-\infty}^{\infty} \left[ \frac{\partial  \bar{\mathcal{R}}}{\partial c_-} \right]_{\rm PL} dx &=& 0, \nonumber \\
\int_{-\infty}^{\infty} \left[ \frac{\partial  \bar{\mathcal{R}}}{\partial \xi_-} \right]_{\rm PL} dx &=&  \frac{8\epsilon}{3}   \frac{a^4 c }{w},\nonumber \\
\int_{-\infty}^{\infty} \left[ \frac{\partial  \bar{\mathcal{R}}}{\partial w_-} \right]_{\rm PL} dx &=& 0, \nonumber \\
\int_{-\infty}^{\infty} \left[ \frac{\partial  \bar{\mathcal{R}}}{\partial \phi_-} \right]_{\rm PL} dx &=&  -\frac{8\epsilon}{3}   \frac{a^4}{w}.\nonumber
\end{eqnarray}
%
By combining conservative and non-conservative contribution, the NCVA yields
the following equations of motion for the NLS with density dependent loss:
\begin{equation}
\begin{cases}
  \dot{a} = - \frac{2}{3} \epsilon a^3   - a b - \frac{2}{\pi^2} \epsilon a^3, \\[1.0ex]
\dot{b}  = \frac{2}{\pi^2}w^4 - \frac{2}{\pi^2}a^2 w^2 - 2 b^2, \\[1.0ex]
\dot{c} = 0 , \\[1.0ex]
\dot{\xi} = c, \\[1.0ex]
\dot{w} = -2bw - \frac{4}{\pi^2} \epsilon a^2 w, \\[1.0ex]
\dot{\phi} = \frac{5}{6} a^2 - \frac{1}{3} w^2 + \frac{1}{2} c^2, \end{cases}
 \label{eq:NVCADD}
\end{equation}
which correspond to the same dynamics as the conservative case (\ref{eq:CODEsNLS}) with the
added nonlinear loss terms $-(2/3 + 2/\pi^2 )\epsilon a^3$ for the evolution of the amplitude
and $-4/\pi^2 \epsilon a^2 w$ for the evolution of the inverse width.  The resulting dynamics
for $\epsilon=0.1$ are depicted in Fig.~\ref{fig3}. We note that even larger values
of the nonlinear loss (cf.~$\epsilon=1$) yield similar results.
As for the linear loss case, we find very good agreement between the full NLS
dynamics and the NCVA results despite the fact that the loss is in this case
nonlinear. While the main properties of the solution dynamics (e.g. the
decrease of the amplitude and increase of the width) persist, there
are also nontrivial differences from the case of the previous subsection
(e.g. in that the decay of the amplitude here follows a power law
rather than an exponential).

\subsection{Exciton-polariton condensates}


The third, and final, example that we consider
in the present work stems from the realm of exciton-polariton condensates.
In this context, the condensing ``entities'' are excitons, namely
bound electron-hole pairs. When confined in quantum wells placed in
high-finesse microcavities, these excitons develop strong coupling
with light, forming exciton-photon mixed quasi-particles known
as polaritons \cite{RMP}.
An especially interesting feature of such polariton condensates is that their finite temperature
leads the polaritons to possess a finite lifetime ---they can only exist for a few picoseconds
in the cavity before they decay into photons.
Hence, in this case, thermal equilibrium
can never be achieved and the system produces a genuinely
far-from-equilibrium condensate in which external pumping from a reservoir
of excitons counters the loss of polaritons.
Exciton-polariton condensates offer numerous key features of the superfluid 
character for exciton-polariton condensates including:
the flow without scattering (analog of the flow without friction) \cite{amo1},
the existence of vortices \cite{lagou1}
and their interactions \cite{roumpos,roumpos2},
the collective superfluid dynamics \cite{amo2},
as well as remarkable applications such as spin switches \cite{amo3},
and light emitting diodes \cite{amo4} operating even near room temperatures.

The pumping and damping mechanisms associated with polaritons enable
the formulation of different types of models. One of these,
proposed in Refs.~\cite{berloff1,kbb,berl_review},
suggests the use of a {\em single} NLS-type equation for the polariton
condensate wavefunction which incorporates the above gain-loss mechanisms.
Specifically, this model, based on a repulsive ($g=-1$)
NLS equation with linear gain ($i\chi(x) u$)
and density dependent loss ($-i\sigma |u|^2 u$) terms, can be written
in the following non-dimensional form~\cite{ref8,berloff1}:
\begin{equation}
iu_t + \frac{1}{2} u_{xx} - |u|^2 u - V(x)u =  i\left[\chi(x) - \sigma  |u|^2\right] u, \label{eq:NLSP}
\end{equation}
where $\sigma$ is the strength of the density dependent loss and
we consider the localized, spatially dependent, gain
\begin{equation}
\chi(x) = \alpha\exp\left(-\frac{x^2}{2 \beta^2}\right),
\label{eq:gainP}
\end{equation}
induced by a
laser pump of amplitude $\alpha$ and width $\beta$.  In the model, we use a  general quadratic potential of the form
\begin{equation}
V(x) = \frac{1}{2} \Omega^2 x^2.
\label{eq:potP}
\end{equation}
\begin{figure*}[t]
\centering
\includegraphics[width=8.9cm]{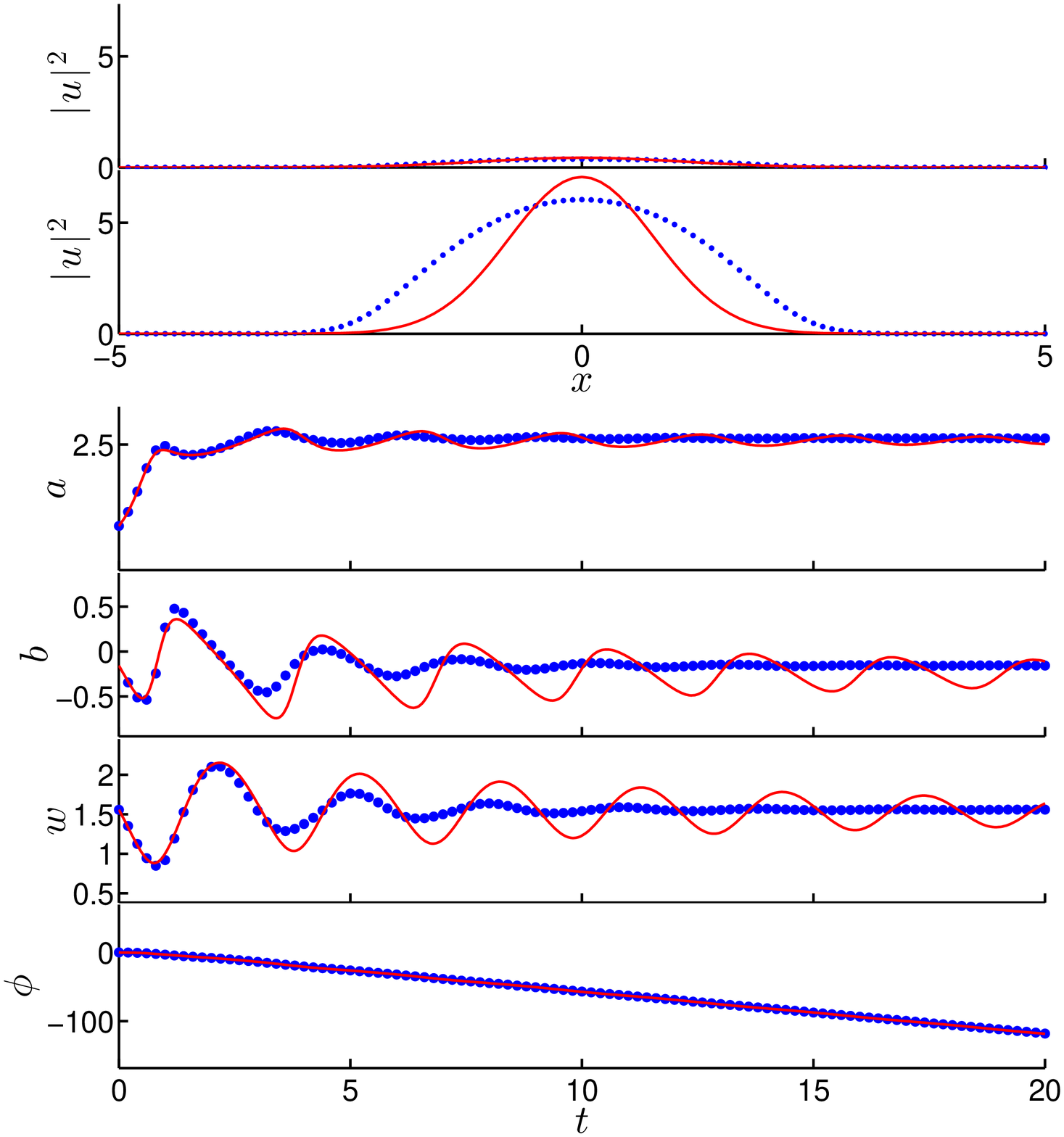}
\includegraphics[width=8.9cm]{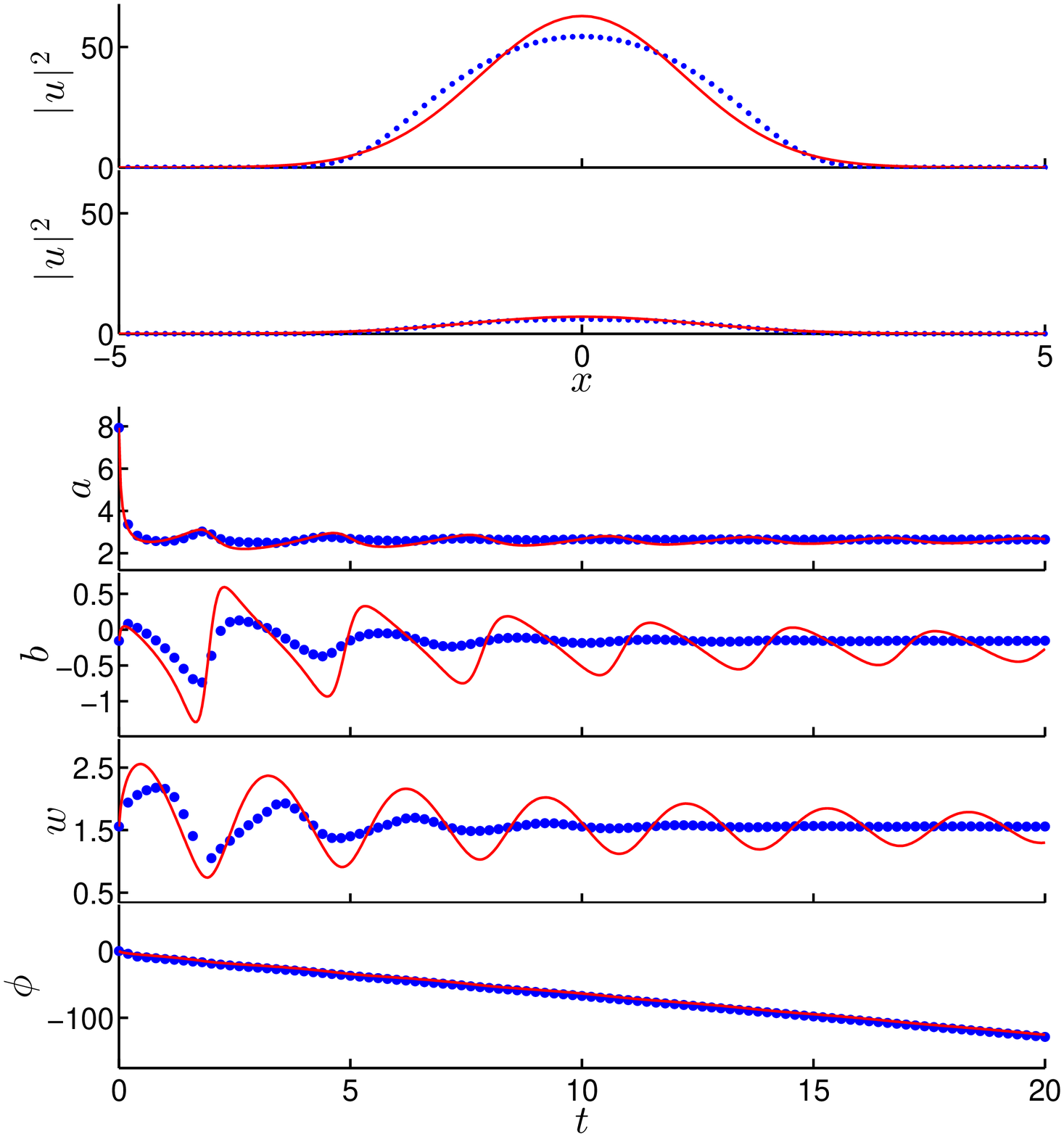}
\vspace{-0.2cm}
\caption{
Evolution of the ground state of Eq.~(\ref{eq:NLSP}) 
in the presence of a linear spatially
dependent gain (\ref{eq:gainP}) with
$\alpha =2$ and $\beta=2$, and density dependent loss of strength 
$\sigma =0.37$, as well as a harmonic potential (\ref{eq:potP}) 
of strength $\Omega = \sqrt{2}$.
To craft initial conditions with amplitudes below (left set of panels) and
above the equilibrium amplitudes we first computed the
steady state of the NLS (\ref{eq:NLSP}) which,
after projection, using least-squares fitting, into the Gaussian ansatz 
(\ref{eq:GaussAnsatz}) yields the following initial parameters:
amplitude: $a(0) = a_e \equiv 2.6431$ (equilibrium), width:  $w(0)=1.5583$, 
chirp: $b(0) = -0.1563$, and phase: $\phi(0)=0.2415$.
Then, we used, for the left set of panels, an initial condition with
$a(0)= 0.6608 = a_e/4$, i.e., four times {\em smaller} than the equilibrium
solution; while for the the right set of panels we used an initial
condition three times {\em larger}  than the equilibrium solution,
i.e., $a(0)= 7.9292 = 3 a_e$.
Depicted are the comparison of the NCVA approximation of Eq.~(\ref{eq:NCVAP})
(red lines) with the full, numerical, NLS evolution of Eq.~(\ref{eq:NLSP})
(blue dots).
The top two panels depict the density $|u|^2$ at the initial time (top subpanel)
and at time $t=50$ (second subpanel).
The bottom four subpanels depict the evolution of the NCVA ansatz parameters
$a$, $b$, $w$, and $\phi$. For the full NLS evolution the parameters are
extracted by projecting the current solution into the NCVA ansatz using 
least-squares fitting.
\label{fig4}}
\end{figure*}

%

To apply the NCVA, we take again the $\bar{u}_1= u_A(x,t; \vec{p_1})$ and $\bar{u}_2 = u_A(x,t;\vec{p_2})$ ans{\"a}tze now defined as a Gaussian of the form
\begin{equation}
u_A (x,t; \vec{p}) = a e^{ -\frac{x^2}{2w^2}}  e^{i\left(b x^2 + \phi\right)}, \label{eq:GaussAnsatz}
\end{equation}
where the ansatz parameter $\vec{p}_i = (a_i, w_i, b_i, \phi_i)$ for $i=1$ and 2
represent, respectively, the amplitude, width, chirp and phase of the ansatz solution. It should be clarified here that our aim in this case 
(and in selecting this particular ansatz) is to 
characterize the breathing motion of a ground state inside the trap,
rather than to characterize the translational dynamics of the wavefunction
(the latter would require a different ansatz, parametrizing the wavefunction
also by its center position).
According to the NCVA method the Lagrangian is $\bar{\mathcal{L}} = \bar{\mathcal{L}}_1 - \bar{\mathcal{L}}_2 + \bar{\mathcal{R}}$, where the conservative parts of the
Lagrangian correspond to
\begin{eqnarray}
\bar{\mathcal{L}_1} &= \frac{i}{2} \left(\bar{u}_1 \bar{u}_{1,t}^* - 
\bar{u}_1^* \bar{u}_{1,t}\right) &+ \frac{1}{2} |\bar{u}_{1,x}|^2 + \frac{1}{2}|\bar{u}_1|^4 \quad \nonumber \\[1.0ex]
 & &+ V(x) |\bar{u}_1|^2,  \quad\\[1.0ex]
\bar{\mathcal{L}}_2 &= \frac{i}{2} \left(\bar{u}_2 \bar{u}_{2,t}^* - \bar{u}_2^* \bar{u}_{2,t}\right) &+ \frac{1}{2} |\bar{u}_{2,x}|^2 + \frac{1}{2}|\bar{u}_2|^4  \quad \nonumber \\[1.0ex]
& &+ V(x) |\bar{u}_2|^2, 
\end{eqnarray}
and $\bar{\mathcal{R}}$ has the same type of density dependent loss 
and a linear gain (equivalent to the negative of linear loss)  
shown in the previous test cases.  The non-conservative terms are defined as follows:
\begin{eqnarray}
\bar{\mathcal{R} }&=& \, \bar{\mathcal{P}} u_-^*+ \bar{\mathcal{P}}^*u_-,  \\
&=& -i \chi(x) \left( \bar{u}_2 \bar{u}_1^* -  \bar{u}_1  \bar{u}_2^* \right) + i\sigma [ | \bar{u}_1|^2 \left(  \bar{u}_2  \bar{u}_1^* -  \bar{u}_2^*  \bar{u}_1 \right)  \nonumber \\
&&  +  | \bar{u}_2|^2 \left(  \bar{u}_2  \bar{u}_1^* -  \bar{u}_2^*  \bar{u}_1 \right) +  \bar{u}_2  \bar{u}_2  \bar{u}_1^*  \bar{u}_1^* -  \bar{u}_1  \bar{u}_1  \bar{u}_2^*  \bar{u}_2^2  ]. \nonumber
\end{eqnarray}
For all the parameters we made the substitutions of $\pm$ coordinates into the expression for the total Lagrangian and from the $\bar{\mathcal{L}}_1$ and $\bar{\mathcal{L}}_2$ parts we recover the conservative Euler-Lagrange equations for a Gaussian ansatz with four parameters.
From the non-conservative term $R$, we expand in the $\pm$ coordinate systems and find the integrals, which are combinations of the integrals for linear gain and density dependent loss: 
\begin{eqnarray}
\int_{-\infty}^{\infty} \left[ \frac{\partial  \bar{\mathcal{R}}}{\partial a_-} \right]_{\rm PL} dx &=& 0,  \nonumber  \\
\int_{-\infty}^{\infty} \left[ \frac{\partial  \bar{\mathcal{R}}}{\partial b_-} \right]_{\rm PL} dx &=& -\frac{\sqrt{2\pi}}{4}  \sigma a^4 w^3 + \frac{2\sqrt{2\pi} \alpha \beta^3 a^2 w^3}{\left( w^2 + 2\beta^2 \right)^{3/2}} , \nonumber \\
\int_{-\infty}^{\infty} \left[ \frac{\partial  \bar{\mathcal{R}}}{\partial w_-} \right]_{\rm PL} dx &=& 0, \nonumber \\
\int_{-\infty}^{\infty} \left[ \frac{\partial  \bar{\mathcal{R}}}{\partial \phi_-} \right]_{\rm PL} dx &=& -\sqrt{2\pi} \sigma a^4 w + \frac{2\sqrt{2\pi} \alpha \beta a^2 w}{\sqrt{w^2 + 2 \beta^2}}  .\nonumber
\end{eqnarray}
Finally, combining non-conservative and conservative terms, the NCVA yields
the approximate 
equations of (breathing) motion for the exciton-polariton 
ground-state condensate of the form:
 \begin{equation}
 \begin{cases}
  \dot{a} = \frac{\sqrt{2}}{8} \sigma a^3 -\frac{3\sqrt{2}}{4} \frac{\sigma a^3 w^2}{w^2 + 2\beta^2 } + \frac{3\sqrt{2}}{2} \frac{\alpha \beta a w^2}{\left( w^2 + 2 \beta^2 \right)^{3/2}} \\[1.0ex]
\quad \quad \quad   -\frac{3\sqrt{2}}{2} \frac{\sigma \beta^2 a^3 }{w^2 + 2\beta^2} + \frac{2\sqrt{2} \alpha \beta^3 a}{\left( w^2 + 2\beta^2\right)^{3/2}} - ab , \\[2.0ex]
\dot{b}  = \frac{\sqrt{2}}{4} \frac{a^2}{w^2} + \frac{1}{2 w^4} - \frac{1}{2} \Omega^2 - 2b^2, \\[2.0ex]
\dot{w} = -\frac{5\sqrt{2}}{4} \sigma a^2 w + \frac{3\sqrt{2}}{2} \frac{\sigma a^2 w^3}{w^2 + 2\beta^2} - \frac{\sqrt{2} \alpha \beta w^3}{\left( w^2 + 2\beta^2 \right)^{3/2}}  \\[1.0ex]
\quad \quad \quad  +\frac{3\sqrt{2} \sigma \beta^2 a^2 w }{w^2 + 2\beta^2} + 2 w b, \\[2.0ex]
\dot{\phi} = -\frac{5\sqrt{2}}{8} a^2 -\frac{1}{2w^2} .
\end{cases}
 \label{eq:NCVAP}
 \end{equation}
%

The comparison between the NCVA ODEs and the numerically integrated NLS is depicted in 
Fig.~\ref{fig4} for initial conditions
below (left set of panels) and above (right set of panels) the equilibrium
for the NLS.
Below and above equilibrium refers, respectively, to initial solution
amplitudes below and above those theoretically predicted to be at
equilibrium. 
In the figure we use the coefficients $\sigma = 0.37$, $\alpha =2$, $\beta=2$, 
and $\Omega = \sqrt{2}$ which ensures that the state with no excitation 
(i.e., without a dark soliton) is stable (see Ref.~\cite{ref8}). 
The top two panels show spatial profiles of the densities
$|u|^2$ at the initial time ($t=0$) and a final time of $t=50$
for the NLS and NCVA solutions. The remaining panels depict the evolution
of the NCVA parameters.
As before, to compare the NCVA evolution of the parameters to the full 
NLS numerics, the numerical NLS solutions are projected (using least-squares fitting) 
onto the variational ansatz $u_A$ at discrete time intervals in order extract 
the parameters $\vec{p}$.
As observed from the figure, the agreement between the NCVA system of 
ODEs and the numerical PDE integration is good, although of lower
quality in comparison to the rest of our considered cases.  
%
%
%
From Fig.~\ref{fig4} it is clear that both the 
original NLS dynamics and its approximation using the NCVA are in very good
qualitative agreement and good quantitative agreement. The lack of a better
quantitative agreement stems from the fact that the solution to the original
NLS problem is only approximately a Gaussian ---cf.~configuration discrepancy 
between the converged states in the the second panel of the left set of panels
in Fig.~\ref{fig4}.
Actually, the solution is only
close to a Gaussian for small atom number, while upon increasing
the atom number it approaches the so-called Thomas-Fermi
(inverted parabola) profile~\footnote{In principle one
could use a better suited ansatz like the $q$-Gaussian proposed in 
Ref.~\cite{qGauss} at the expense of obtaining more complicated reduced
NCVA ODEs.}
In all cases (below or above the stationary steady state), the dynamics
of the NCVA and NLS converge (in an oscillatory manner) to their respective 
stable solutions and do so consistently with respect to each other.

\section{Conclusions \& Future Challenges}
\label{sec:conclu}

In this work we have extended the non-conservative variational formulation 
recently
proposed by Galley~\cite{ref1}. This was 
originally developed for classical mechanics,
namely for systems with few degrees of freedom (and subsequently extended,
including in the form of a variational approximation, to nonlinear
Klein-Gordon models in~\cite{ref4}), to the nonlinear Schr\"{o}dinger
equation (a complex partial differential equation, i.e., infinite number
of degrees of freedom).
By using this non-conservative approach on a suitably chosen ansatz,
it is possible to reduce the original infinite-dimensional dynamics to
a system of ordinary differential equations on the ansatz parameters.
We show that the resulting non-conservative variational method for the
nonlinear Schr\"{o}dinger equation is equivalent to the (linear) perturbative
variational method.
We provide several examples to test the validity of the non-conservative 
variational approach.
In particular, we include examples with linear and density-dependent (nonlinear)
loss. We also showcase the application of this method to exciton-polariton
condensates that are inherently lossy and need a pumping term to balance losses.
In all cases we see a very good qualitative and also,
in principle, quantitative agreement (with the partial
exception of the exciton-polariton condensate) between
the original dynamics and statics for the nonlinear Schr\"{o}dinger equation
and its corresponding reduced non-conservative variational counterpart.
This agreement seems to be preserved even when the non-conservative terms
are of the same order of magnitude as the other (conservative) terms. This
is in contrast with perturbation methods that intrinsically rely on the
non-conserved terms being small when compared to the conserved ones.

It would be interesting to apply in more detail the non-conservative 
variational methodology studied here further 
to exciton-polariton condensates ---more specifically---
that are, intrinsically, open systems of high interest and significant
impact to ongoing experimental efforts. In particular, such an approach 
would be valuable towards detecting 
the boundaries for stability inversion reported 
in Ref.~\cite{ref8}. It was noted in that work that, surprisingly, 
for some parameters values, the (originally) ``excited'' dark soliton 
state (with one nodal point) becomes stable in favor of 
the nodeless cloud state (the state without the dark soliton corresponding
to the original ground state of the system in the absence on non-conservative
terms) which in turn loses its stability. A systematic study of the
dark soliton and its stability in such condensates can be found
in Ref.~\cite{smirnov}.
Also valuable would
be to extend this approach to the two-dimensional case where the equivalent 
stability inversion for vortices and rotating lattices has also
been reported~\cite{berloff1}.

This work opens, more broadly,
a few interesting avenues for future explorations.
In particular, it opens the possibility to apply a similar methodology
to other models described by partial differential equations which
contain non-conservative terms. The non-conservative variational
methodology could prove very useful in cases where traditional perturbative
techniques, relying on the smallness of the non-conserved terms, fail
due to the magnitude of the perturbations.
One such model is the quintessential complex Ginzburg-Landau
equation~\cite{Aranson:02} that models an extremely wide
range of open systems including nonlinear waves, phase transitions,
superconductors and superfluids, among others.

\vspace{2mm}

{\it Acknowledgements.} P.G.K.~gratefully acknowledges the support of
NSF-DMS-1312856, BSF-2010239, as well as from
the US-AFOSR under grant FA9550-12-1-0332,
and the ERC under FP7, Marie Curie Actions, People,
International Research Staff Exchange Scheme (IRSES-605096).
The work of PGK at Los Alamos is partially supported
by the US Department of Energy.
PGK would also like to thank S. Ch\'{a}vez Cerda
for a useful discussion on this topic, as well as for
pointing us to the relevant recent work of Ref.~\cite{ref3}.

\end{document}